\documentclass{acmart}
\AtBeginDocument{%
  \providecommand\BibTeX{{%
    \normalfont B\kern-0.5em{\scshape i\kern-0.25em b}\kern-0.8em\TeX}}}

\setcopyright{acmcopyright}
\copyrightyear{2021}
\acmYear{2021}
\acmDOI{}
\acmSubmissionID{}

\acmConference[MAB 20]{Media Architecture Biennale 2020}{June 28th-July 2th 2021}{Amsterdam,Utrecht, NL}
\acmBooktitle{}
\acmPrice{}
\acmISBN{}

\usepackage{subcaption}
\usepackage{microtype}

\begin{document}
\title{Distancing interfaces—improvisational media architectures for place-based discourse under lockdown}

\author{Dietmar Offenhuber}
\email{d.offenhuber@northeastern.edu}
\orcid{0000-0002-8434-7786}
\affiliation{
  \institution{Northeastern University}
  \streetaddress{360 Huntington Ave.}
  \city{Boston}
  \state{Massachusetts}
  \country{USA}
  \postcode{02115}
}
\author{Sam Auinger}
\email{samauinger@me.com}
\affiliation{
  \institution{Independent}
  \country{DE}
}

\begin{abstract}
The paper presents the case study of BERLIN\_LOKAL\_ZEIT, a collaborative artistic media project aiming to document and reflect individual experiences of the COVID-19 lockdowns in Berlin, Germany. What began as an observational effort in spring 2020 became a year-long archive, exhibition, and broadcasting platform that generated various hyper-local interfaces. The paper articulates an improvisational approach to media architecture in the form of self-reflection by the project initiators. Necessitated by the limitations imposed by the lockdowns on cultural production and public discourse, the paper presents an alternative conceptual approach to media architecture that is not based on a fully-specified technological infrastructure for discourse and interaction but instead on improvisational practices that manifest themselves in different technical interfaces. This improvisational approach is not merely a mode of production, but raises questions of about the discourse of media architecture and its underlying assumptions of methodological rigor. In traditional HCI, systems and artifacts are often presented as stable and fully specified, while their users are considered interchangeable. In contrast, we consider urban media interfaces as improvisation-driven infrastructures (or \textit{improstructures}) in which the actors and their relationships are stable, while the technologies they create are provisional and in flux. As a consequence, the defining property of media architecture is no longer the structural integration of media into architecture but the embodiment of mediated communication in objects and performances at concrete places. In contrast to familiar concepts such as probes or prototypes \cite{gaver2004, paulos2005}, which may also involve bricolage and invite improvisation, improstructure emphasizes the infrastructural processes of maintaining, tweaking, and extending rather than the artifact and its use. In an environment where many physical spaces are longer accessible and human contact among strangers diminished, the project explores the restorative function and capacities of media architecture for shared reflection and discourse in physical public space.

\end{abstract}


\maketitle
\section{Introduction}
The lockdowns enacted in response to the COVID-19 pandemic have disrupted urban life in many cities of the world. Apart from the hardships that many people experience, the city under lockdown has become unfamiliar, both pleasantly and disturbingly so. As traffic, commerce, and public life retreat, new phenomena emerge. Some of them have become part of the collective discourse: the eerie quality of empty urban spaces, the lack of background noise, or animals reclaiming urban space. Other phenomena, however, present themselves to each individual differently. The collective reflection of such experiences is complicated by the fact that public discourse and interactions have shifted into the digital realm, where facts seem to become more fluid and contested. Using the ongoing art project BERLIN\_LOKAL\_ZEIT \footnote{literally translated, Berlin Local Time, see \url{http://berlinlokalzeit.de}} as a case study, we examine how the discourse in physical public space can be reclaimed under lockdown conditions to promote shared reflection, coping, and healing. We explore how media architectural practices can support this goal and consider which qualities media interfaces should have in order to fulfill this role.  

The BERLIN\_LOKAL\_ZEIT project was initiated by the authors during the first lockdown in Berlin, Germany in spring 2020 with the goal to examine the pandemic from a personal perspective. Invited by the authors, 21 individuals documented the impacts of the lockdown on the experience of the city along sensory and affective dimensions. Despite the personal focus, the project was not intended to be introspective but rather an effort to document and engage external phenomena and events through observation and intervention. After running for almost a year and transforming into an exhibition and performance platform, the project gained an important restorative purpose in a situation where all artistic and cultural events were suspended. Focusing on the affordances of physical space when everything else has moved online, the project explored the capacity of art and culture to connect people living in difficult circumstances and engage them in public dialogue.

Since the lockdown measures prevented a traditional exhibition, the collective had to explore new strategies to mediate between physical and online spaces. The artifacts and infrastructures that resulted from this effort are technologically unsophisticated, yet they constitute media architecture in its most elementary sense. At the same time, the project breaks with tacit assumptions about media architecture---for example, that interfaces and infrastructures enable public discourse but typically do not themselves emerge from it or become its subject. BERLIN\_LOKAL\_ZEIT required a different approach—an exchange on the same level where every aspect of the project is renegotiable, including the technologies and interfaces. The technological artifacts, including a QR code facade, a mobile radio station in public space, sound installations, static and wearable displays, emerged from an improvisational process and take advantage of tools and everyday communication practices. The interaction takes place in multiple channels and modalities—online and offline, synchronous and asynchronous, but always grounded in a particular place. 

While the pandemic archive would deserve separate attention, this paper focuses on the various interfaces and media strategies developed for the subsequent exhibition of the project, which coincided with the second wave lockdown of the pandemic. In the spirit of the project’s epistemic goals, we did not opt for the standardized evaluation methods customary in HCI nor follow a literature-driven taxonomic approach. Instead, the paper employs a qualitative method based on a wealth of collected materials, the authors' self-reflections, conversations and interviews with the project participants. The approach can be compared to the method of auto-ethnography \cite{chang2016,lucero2018}, emphasizing reflection of own life experiences, their contextualization, and analysis in a detailed narrative account. The main difference to BERLIN\_LOKAL\_ZEIT is that in our case, it is the project that continuously documented itself—through its participants’ reflections and interactions with the public in interviews, podcasts, and online discussions. Our goal is to examine the dynamics of collaborative production as a way coping with a difficult situation, rather than claiming generalizable results. We nevertheless think that the project holds important lessons for media architecture and invites re-examination of some of its underlying assumptions. We will discuss and summarize practical lessons at the end of the paper.

\begin{figure}
    \centering
    \includegraphics[width=\linewidth]{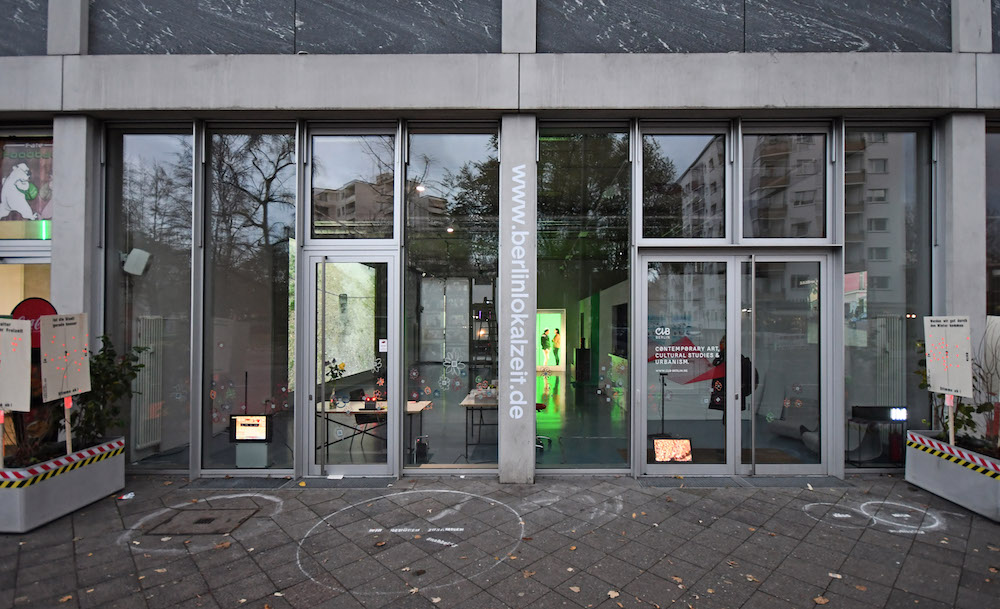}
    \caption{View of the CLB facade from Moritzplatz, shortly after the opening in Nov 20.}
    \label{fig:clb}
\end{figure}

\section{Scope and Data Sources}
This paper draws from a range of data collections. The pandemic archive was created in a private Slack workspace in April 2020 and remains active up to the present in February 2021. 21 individuals recorded their observations in the form of text, image, video, and audio recordings. At the end of February 2021, the archive included 1331 entries, including 52 audio recordings in the form of spoken observations and field recordings, 627 images, 33 movies, and 619 text entries. For the launch of the exhibition on November 20, we began collecting material also on a dedicated website, which included a browser and a visualization for the pandemic archive and, as of February 2021, documented 10 site-specific interventions, 7 sound walks, and 75 radio features. After the launch of the website, a large part of the material collection and discourse moved to a public Telegram channel associated with the project.\footnote{See \url{https://berlinlokalzeit.de/page/live/}} The radio platform Radio Aporee, a project by the participant and sound artist Udo Noll, became another major component of the project, providing mobile radio broadcasts, streams, and an archive of auditory contributions.\footnote{See \url{https://aporee.org}} We also conducted public, 18 non-directed interviews with the participants, published on Radio Aporee and the project website.\footnote{See \url{https://aporee.org/t/berlinlokalzeit/list.php}} 

\section{Media Architecture as Improstructure}

Media facades that “use [...] digital media to transform the outer shell of a building” \cite{haeusler2009} are typically designed as a permanent part of its infrastructure, for example, as a building-sized screen. This permanence is often emphasized to underscore the tight coupling between digital and physical spaces. As such, media facades share similarities with an operating system of a computer: strictly specified in terms of format and structure, but otherwise open to run arbitrary media content; advertisements, public messages, or art. Media facades are thus not only technical- but also civic platforms that enable individual expression in public space—for example, by displaying art and poetry, statements and opinions from citizens. Recent media architecture discourse has started to look beyond such fixed installations, giving considerable attention to temporary, pop-up, hybrid interventions in public space that often appropriate ubiquitous technologies such as smartphones \cite{caldwell2014,caldwell2017}. It may be tempting to put such low-tech approaches in contrast with the media infrastructures found on the facades of corporate headquarters. However, it needs to be stressed that both forms of media architecture, planned as part of the architecture and DIY, can act as a coherent infrastructure that offers a range of affordances for interaction, expression, and learning.

Nevertheless, we can make a distinction with regards to how such infrastructures are made and re-made, maintained, and cared for. The term \textit{infrastructuring} implies that planning and production are not limited to a well-defined design phase; they are an ongoing process \cite{dantec2013}. In DIY projects, planning and execution often coincide, making their production a form of organizational improvisation \cite{cunha1999}. The framework of \textit{improstructure} attends to how infrastructures are governed through improvisational processes among actors who do not always share a pre-determined plan. They create and govern systems through a jazz-like process of “call and response,” reacting to each other’s actions and using the resources available to them at the moment \cite{offenhuber2018a}. Another perspective focuses on the users and their improvisational interactions; how they appropriate and creatively abuse their affordances \cite{kloeckl2020}. 

To a varying extent, all infrastructures involve such improvisational aspects. Their evolution and transformations are a frequent theme in the literature on socio-technical systems, from large technological systems such as power grids and the internet to \textit{inverse infrastructures} built and maintained by their users \cite{egyedi2012,hughes1987}. However, the terminologies used to describe such systems often obscure these underlying processes and imply a much simpler state of affairs. The concept of the platform itself serves as an exemplar. In a technical context—as in, for example, the windows platform—it refers to a set of technologies with strictly defined specifications that encompass software, hardware, peripherals, and protocols for interaction. Applied to a media facade, this perspective suggests that the infrastructure is open and allows for uses unforeseen by the platform’s developers. Simultaneously, the specifications also define hard limits such as frame rate or resolution \cite{offenhuber2014b}.

However, the term \textit{platform} has several conflicting meanings and implications, even within a technical context. Management theorist Claudio Ciborra described the platform company—based on the example of the Italian Olivetti corporation—as a shape-shifting entity without a fixed structure that can transform into many different forms as needed \cite{ciborra1996}. While in the computing platform, the fixed technical specifications hold everything together, in Ciborra’s platform the same technical structures are ephemeral and exchangeable while the platform coheres through the formal and informal relationships of its actors. To some extent, this also applies to social media platforms and platform companies like Uber and AirBnB, which evolved with the practices of their users, especially in their early stages. Even computing platforms, where every technical parameter is supposedly defined and specified, often present themselves as an informal bricolage with incompatibilities between components of different origins, undocumented bugs, and makeshift repairs.

The goal in this paper is to expand the perspective on media architecture from the emphasis of structural permanence—expressed in the objective “to use information as an additional construction material in architecture like glass, wood or concrete” \cite{wiethoff2017b}—to a perspective that centers the ad-hoc production of interfaces and temporary interventions as catalysts of public discourse. To this end, our case study emphasizes the relational (i.e. interfaces as artifacts and enactments of human relationships) over the ontological (i.e. which structures exist and what are their affordances) aspects of media architecture. The BERLIN\_LOKAL\_ZEIT project is improvisational by design, or rather by re-design at the moment when it became clear that the usual ways of planning an exhibition, developing the necessary technologies, and interfacing with the public were no longer possible.

\section{History of the project}
In April 2020, we invited around 30 individuals — artists and cultural workers working in precarious arrangements, academics, students, and teenagers — after lockdown measures were imposed in Germany at the end of March. Participants were invited based on the following criteria: they work in the cultural sector, know more than one other participant, and come from all age groups. Rather than selecting a random group, we wanted to include individuals who have an established practice of observing phenomena and translating these observations into creative work. Although most participants did not know each other, we wanted to begin with some prior connections to ease the communication process within the group. 

While media discourse on the pandemic has focused primarily on explaining the abstractions of epidemiological models that justified the lockdown measures, we aimed to offer a Radićally subjective, situated, and phenomenological perspective. The project's premise was that we would take these personal observations seriously and treat them as valuable, reflecting larger societal and political processes. The goal of BERLIN\_LOKAL (the initial name) was therefore not an introspective exploration but the attentive observation and engagement with external phenomena: the changing soundscape, the rhythms of the city, its actors, and the traces they leave. During the first four months of the project, we collected over 600 contributions from around 21 participants, who documented their observations in a private slack workspace in the form of photos, videos, field recordings, texts, and media fragments. A relatively small group created the majority of contributions, documenting their daily observations and the traces of the lockdown in their neighborhoods. Ambivalence stood out as a prevailing sentiment in the early reflections of the collective. No one has clear answers to the many challenges posed by the pandemic or an unambiguous position towards the policies adopted by the city in response.

During this early stage, the changing soundscape was a common theme: the lack of traffic and aircraft noise, which unmasked previously unheard auditory phenomena, unusually silent subway stations and noisy birds, outdoor concerts and parties. A second theme were new traces and reconfigurations of public space: masks as new forms of litter, ubiquitous fences, queues, and distance markers, empty poster walls that are usually filled with event announcements. A third group included bottom-up interventions by citizens, such as pop-up sandboxes on the street, since all playgrounds were closed, and manifestations of protest and rebellion against the measures: graffiti and overpainted signs, monuments dressed up with masks, invitations to illegal disco parties. Many participants, however, went beyond simple documentation, and reflected the implications of these phenomenological changes for concepts of publicness, discourse, and equity. The sound of a helicopter monitoring stay-at-home orders during an otherwise quiet spring day encapsulated the conflicting connotations of the transformed soundscape, from solidarity with the lockdown measures to a slight unease with its implications.\footnote{See \url{https://berlinlokalzeit.de/page/channels.html?c=13&i=5}}

\begin{figure}
    \centering
    \includegraphics[width=\textwidth]{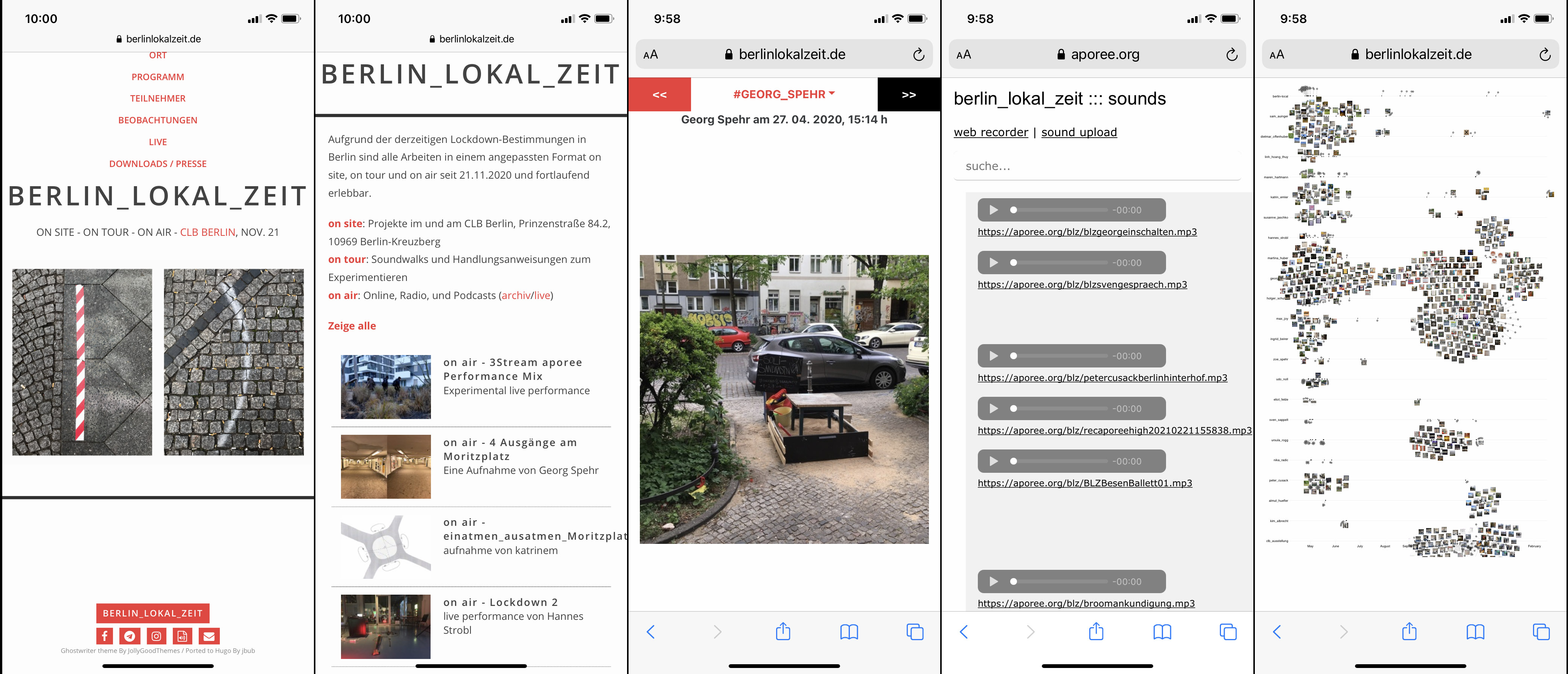}
    \caption{Smartphone Screenshots of the website: homepage, project list, archive view, sound archive, visualization (Kim Albrecht). Feb 21. \url{http://berlinlokalzeit.de}}
    \label{fig:phone}
\end{figure}

\subsection{Phase II — planning an exhibition of the pandemic archive}
As the first lockdown was winding down and the participants' contributions gradually began to dwindle, we started to plan an exhibition of the collected material at the CLB Gallery in Berlin Kreuzberg. For various reasons, the show, initially scheduled for late summer, was postponed until the fall. This new timeline raised our doubts and concerns: with the spring lockdown fading into a distant memory, would people still be interested to hear about these observations? After months of media coverage of the COVID-19 pandemic, how valuable and relevant is the archive as a historical document and a reflection of a recently completed episode? Wouldn't the early explorations of an unknown situation appear hopelessly naive six months later, after everyone has become hardened and slightly cynical by the pandemic?

Re-reading the contributions, we already started to notice a drift in their meaning and significance. What seemed unique and exciting in the beginning has since received so much attention it turned into a cliché. Other issues, initially overlooked, appeared prescient in their anticipation of later controversies, such as those surrounding COVID-19 deniers. These semantic and semiotic drifts reminded us once again why situatedness and personal perspective were so vital for us, as we were less interested in gathering facts rather than probing and diagnosing a situation at a particular moment. In our re-reading, we noticed the importance of temporality. Temporal shifts became the topic of the show, and BERLIN\_LOKAL became BERLIN\_LOKAL\_ZEIT. \textit{Social distancing} is coupled with a regime of \textit{temporal distancing}—from disruptions of schedules to the urban choreographies necessitated by lockdown policies. Before the pandemic, rhythms were highly synchronized, everyone running to the same beat. Now, many participants observed how time has decoupled and individualized, stretched and compressed.

\subsection{Phase III — second lockdown and pandemic interfaces}
By re-conceptualizing the exhibition in this way, we began to plan for an opening in late fall, adding new artworks and performances to the archive installation. However, this plan collided with the second wave of the pandemic. In the week of the planned opening, the city of Berlin was once again placed under lockdown, leading to the ironic situation that the very circumstances that made the project highly relevant again prevented us from opening the exhibition at all.

At this point, media architecture enters the picture. Moving events online has been the typical response in the pandemic, bringing its own advantages and drawbacks. Since the project was about the phenomenology of a concrete place, however, we had to look for different solutions. We accepted the new restrictions as an opportunity to rethink what an exhibition can be, and more generally, how artists can work during the pandemic. Cut off from funding sources and exhibition venues, we saw the need for a more elementary role of art to reflect and respond to what is happening in the world.

Rethinking (once again) the exhibition included questions such as: How can we work with lockdown policies while making them the subject of the exhibition? How can we create meaningful interfaces that stimulate discourse when the said policies and regulations change almost daily? Since traditional strategies for building displays and technologies, which require a significant amount of planning, were not appropriate under the circumstances, we chose an improvisational approach: working with materials and technologies that allow for rapid adaptation and flexible use. Avoiding of the binary choice between creating a physical installation or going completely online, participants crafted a hybrid environment in which digital and physical, synchronous and asynchronous elements are interwoven in complex ways, providing multiple points of interaction. As of February 2021, the site comprises physical installations, a public interface using QR codes to access online resources, coordinate, and participate in performances, soundwalks that can be experienced socially distant together with the artist or alone, mobile radio transmissions and podcasts (Fig. \ref{fig:clb}).

\begin{figure}
    \centering
    \includegraphics[width=0.7\textwidth]{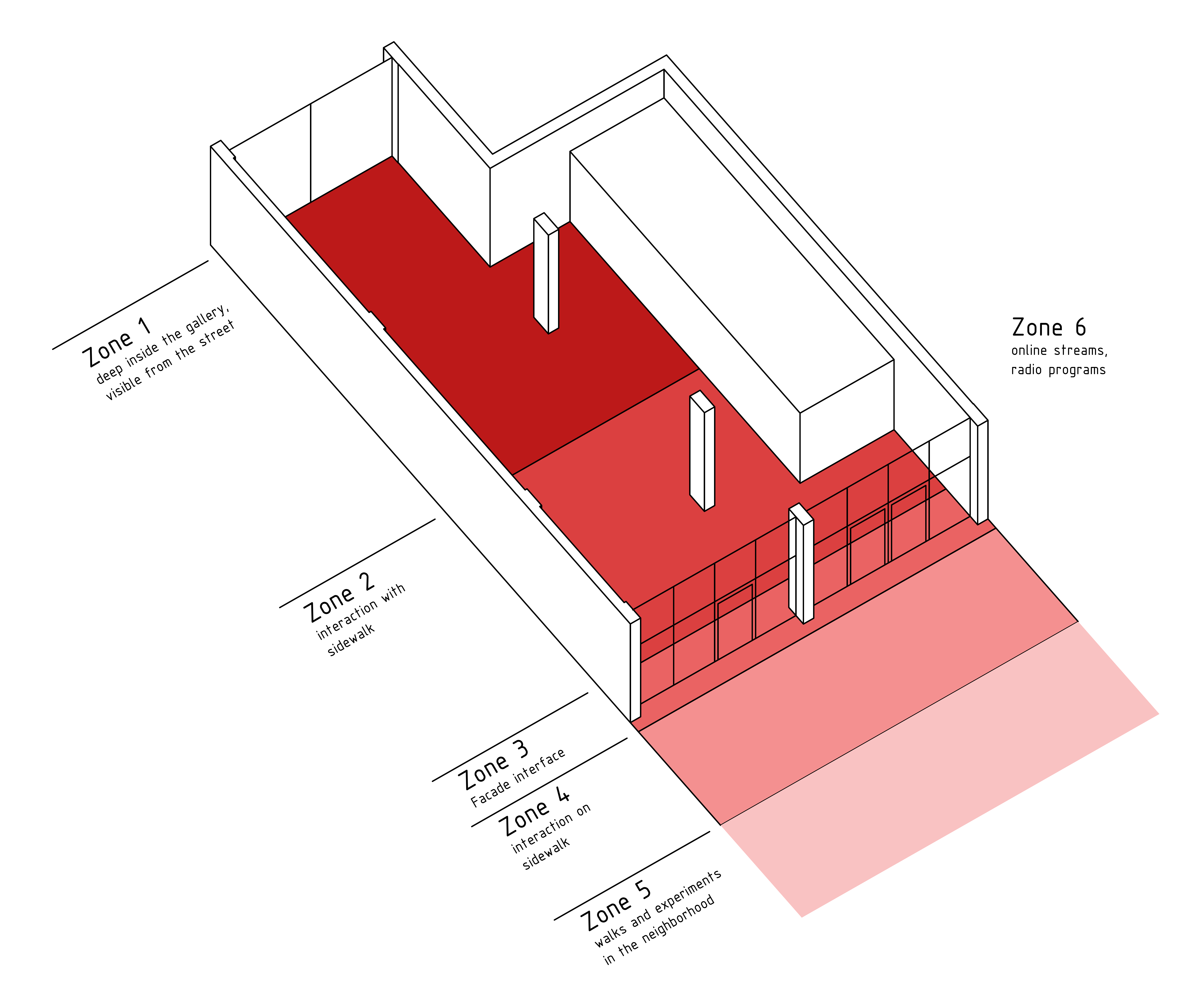}
    \caption{The site of the gallery is divided into six zones, each with a different form of coupling between the physical and online spaces ordered by degree of interaction.}
    \label{fig:zone}
\end{figure}

\section{Semi-permeable interfaces to the public}
In our strategy to intertwine online and physical realms as a \textit{hyperlocal} media architecture\cite{wouters2018a}, the gallery acts as a semi-permeable membrane, an interface divided into several zones \ref{fig:zone}.

As the first zone, the gallery space’s interior is inaccessible due to COVID-19 restriction but is still viewable from the street through the large glass facade. The first measure was to re-orient the artifacts and installations for viewers from the sidewalk. Nika ć projected a doorway into a virtual room where pandemic restrictions do not seem to apply, a video collage of people conversing in intimate proximity. Zoe Spehr’s piece “my window” used the rear window facade to project a matrix of zoom windows recorded during online meetings with her friends. The closed gallery space plays a conceptual role in both pieces: in the case of Radić, by emphasizing the distance and inaccessibility of the pre-pandemic reality, in the case of Spehr, by blending the architectural windows with the virtual windows when viewed from a distance.

\begin{figure}
    \centering
    \begin{subfigure}[t]{0.49\linewidth}
        \centering
        \includegraphics[height=2.1in]{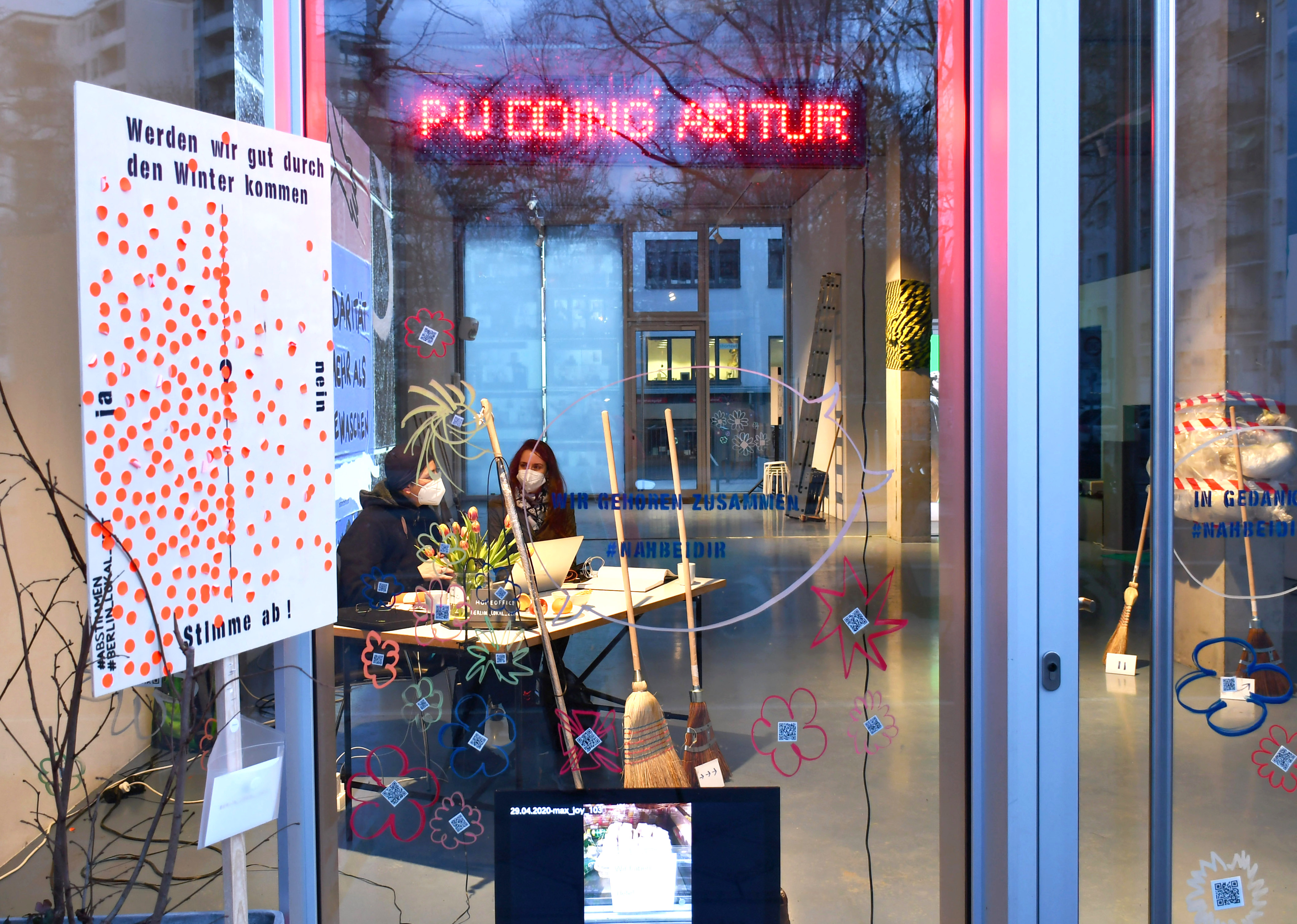}
        \caption{View into the gallery with home office, corona words, voting, and facade interface, Foto Nika Radić, Feb 21.}
        \label{fig:home}
    \end{subfigure}
    ~
    \begin{subfigure}[t]{0.49\linewidth}
        \centering
        \includegraphics[height=2.1in]{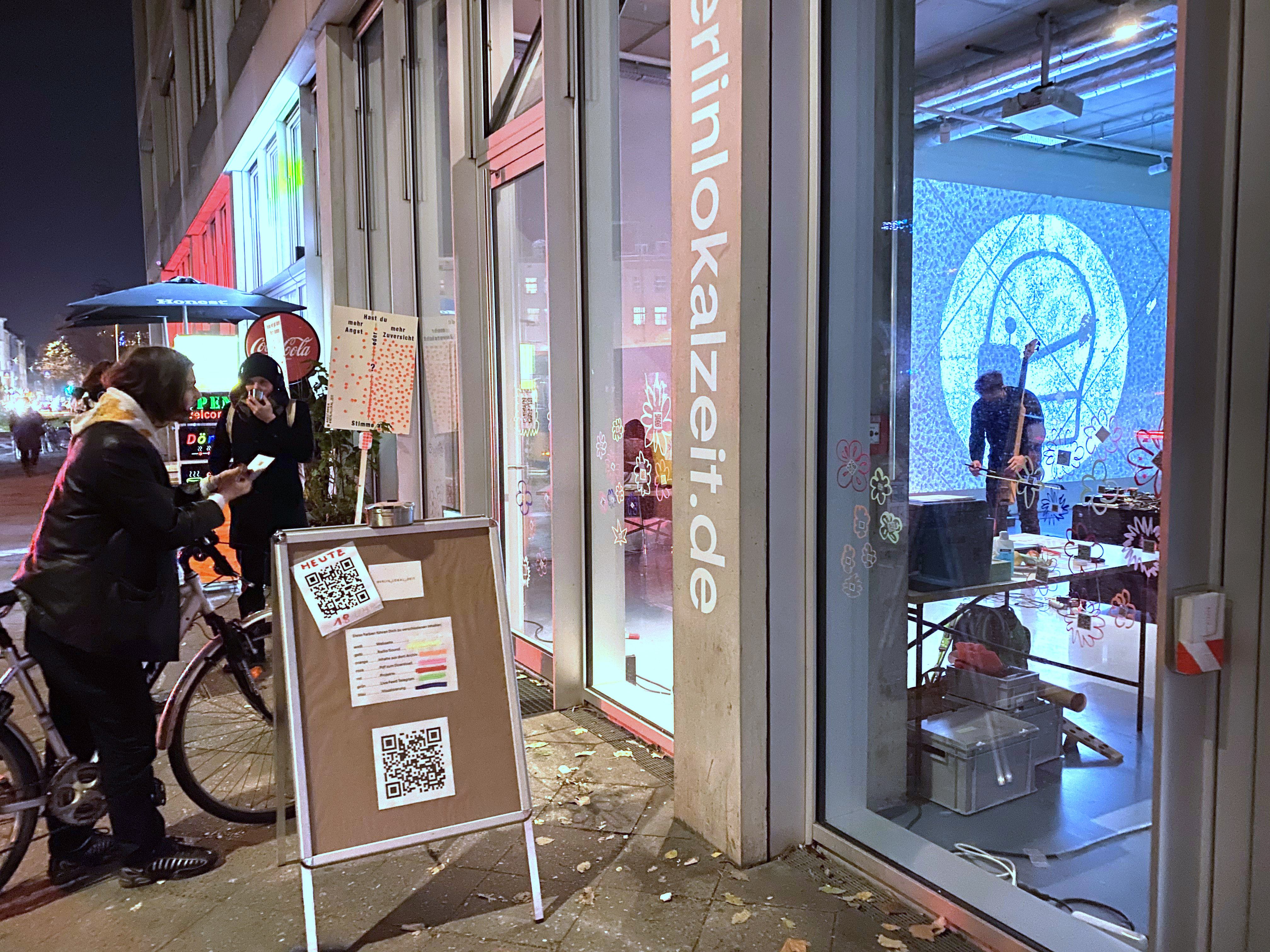}
        \caption{Concert performance Hannes Strobl inside the gallery streamed to the sidewalk, Foto Max Joy, Jan 21}
        \label{fig:hannes}
    \end{subfigure}%
    \caption{Streetside views of the gallery}
\end{figure}

A second zone, still inside the gallery, was located closer to the glass front and allowed for more public engagement. Max Joy’s LED display installation ran a list of “Corona Words,” neologisms he collected from popular media discourse. Videos behind the facade documented recent performances and contributions to the pandemic archive. The installation “home office” offered more interaction, consisting of a shared office desk placed against the gallery’s glass front (Fig. \ref{fig:home}). Participating artists took turns working at this desk during the exhibition, updating their projects and interacting with pedestrians in an ironic reversal of the familiar home office experience. The gallery space was also used for performances; musician Hannes Strobl performed a concert for audiences on the sidewalk, simultaneously streamed online (Fig. \ref{fig:hannes}).

The glass facade itself formed the third zone, a direct interactive interface with the public: QR codes placed directly on the window, surrounded by drawings and annotations, provided access to the project website and archive, acted as a daily updated message board, and offered downloadable instructions for soundwalks, DIY experiments and performances.

\begin{figure}
\begin{subfigure}{0.5\linewidth}
    \centering
    \includegraphics[height=2.15in]{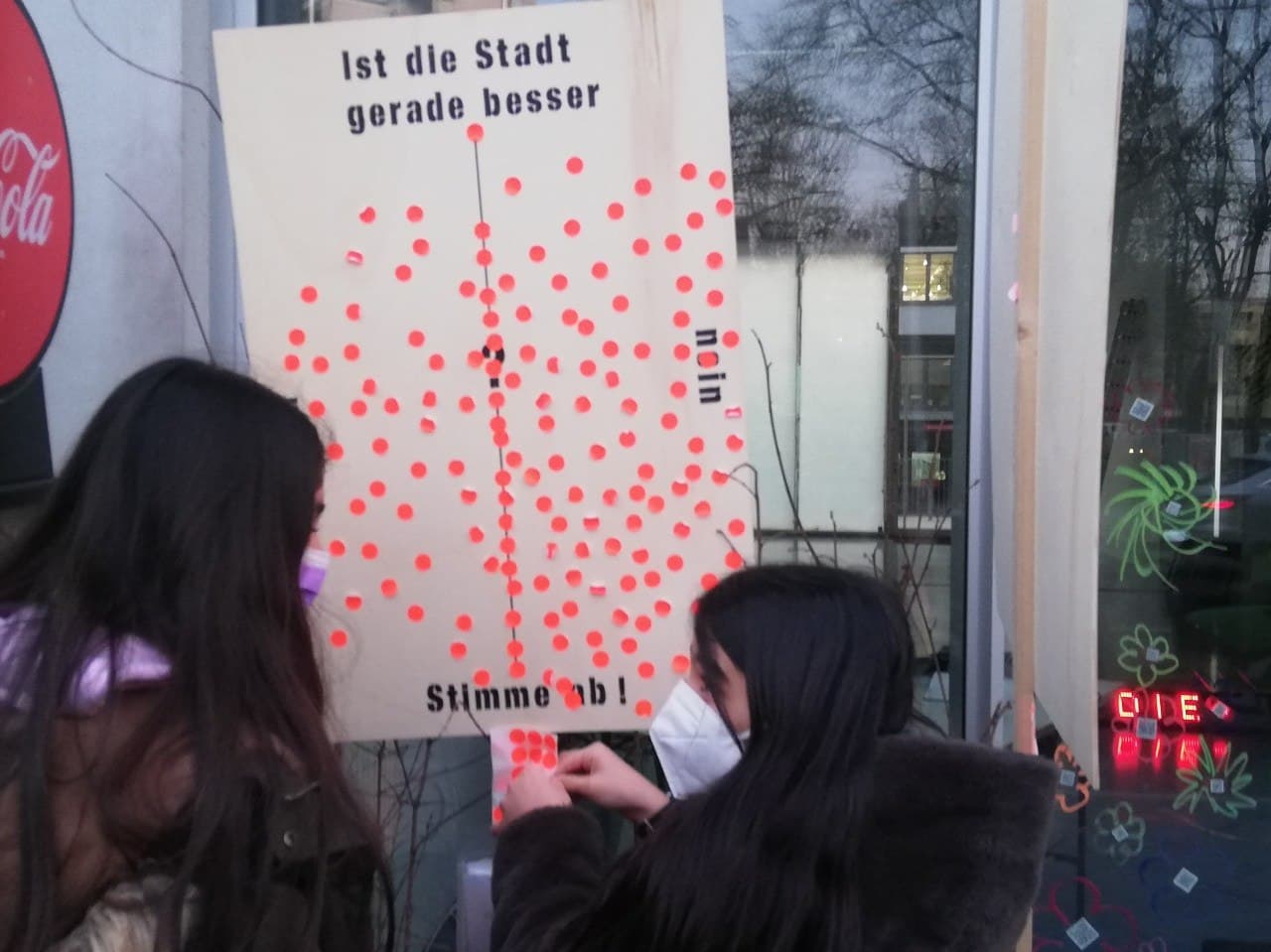}
    \caption{Susanne Jaschko, abstimmen (voting), Nov 20.}
    \label{fig:vote}
\end{subfigure}
~
\begin{subfigure}{0.5\linewidth}
    \centering
    \includegraphics[height=2.15in]{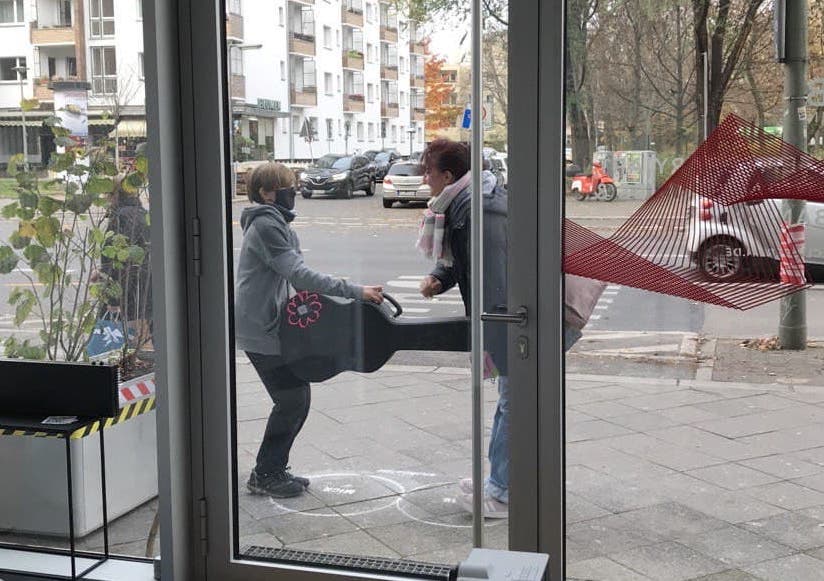}
    \caption{Susanne Jaschko, nah bei dir (close to you) Nov 20.}
    \label{fig:nah}
\end{subfigure}
\label{fig:su}
\end{figure}

A forth zone is located directly in front of the gallery facade. In this zone, Susanne Jaschko installed an installation where pedestrians could vote on questions such as “Is the city now better?” or “Is culture education or leisure?” by placing orange stickers in two zones of a panel (Fig. \ref{fig:vote}). In a second intervention “close to you,” she drew circles on the sidewalk that allow documenting closeness rather than distance, in a play on the ubiquitous distancing markers (Fig. \ref{fig:nah}). Eliot Felde and Zoe Spehr created temporary tattoos based on the iconography of social distancing markers, embodying the mandated behaviors on the skin. The sidewalk was also used for performances and sound installations connected to the exhibition. The gallery also had a second glass facade towards the interior courtyard, which became a space for performances and discourse.

\begin{figure}
    \begin{subfigure}[t]{0.5\linewidth}
        \centering
        \includegraphics[height=2.15in]{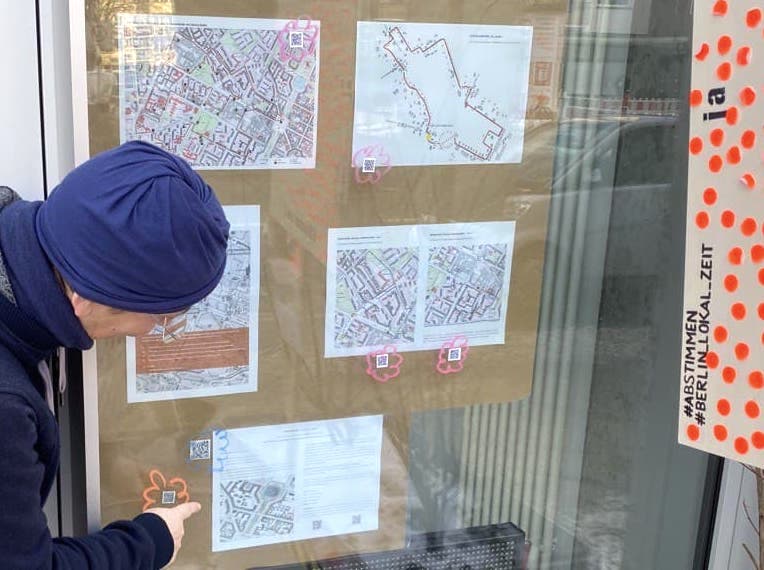}
        \caption{QR-coded soundwalk instructions on the facade}
    \end{subfigure}
    ~
    \begin{subfigure}[t]{0.5\linewidth}
        \centering
        \includegraphics[height=2.15in]{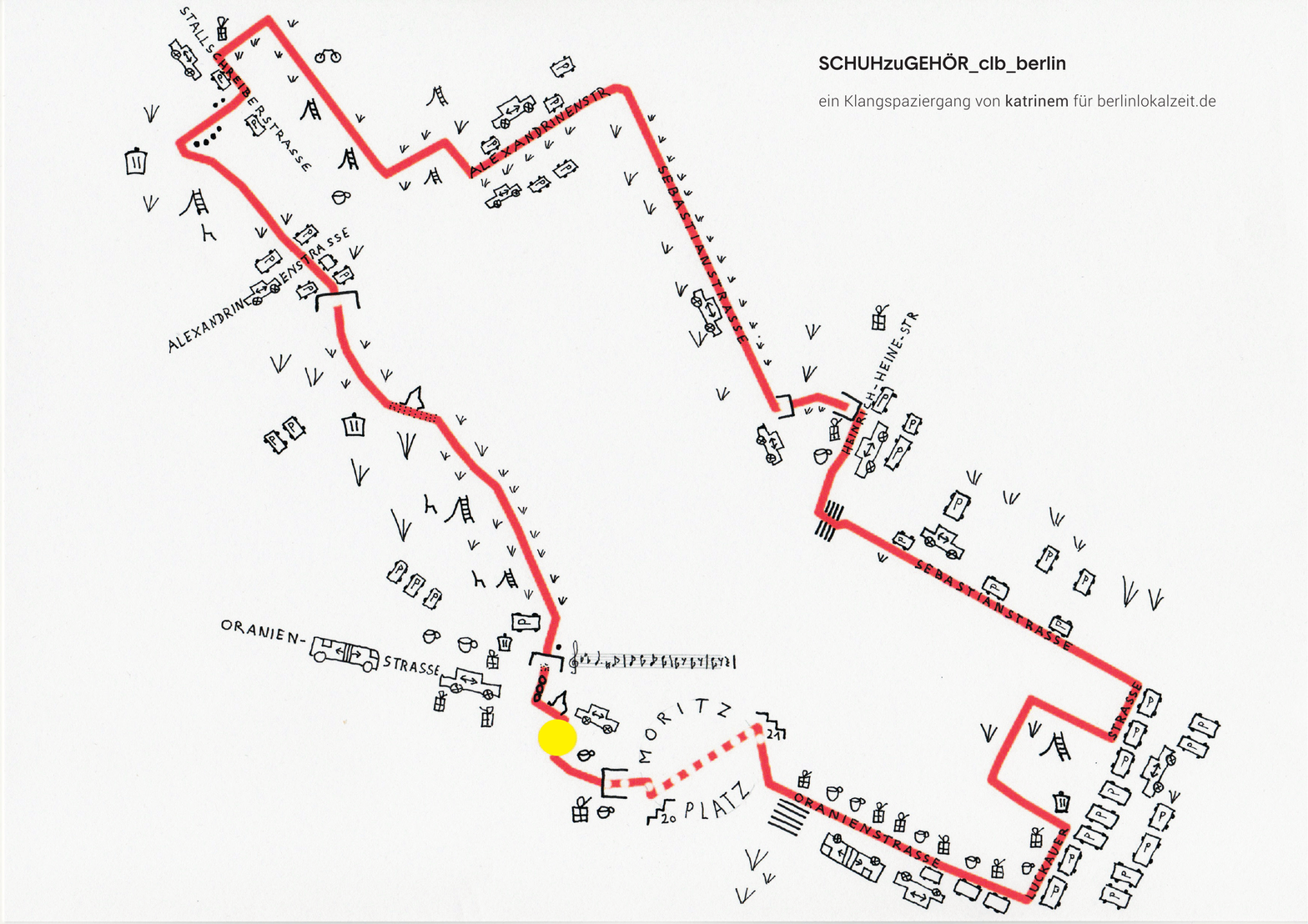}
        \caption{Downloadable instruction for a DIY soundwalk by katrinem (excerpt)}
    \end{subfigure}%
    \label{fig:katrinem}
    \caption{Downloadable instructions for soundwalks on the gallery facade}
\end{figure}

A fifth zone reached far into the adjacent neighborhoods — soundwalks, guided or DIY, invited the public to explore the phenomenological changes of the pandemic, re-evaluate simple acts such as sitting on public benches (Georg Spehr). An example of a DIY experimental protocol are Holger Schulze’s smell exercises, exploring how the ability to smell each other has acquired new connotations. Instructions and maps were offered on the ever-changing facade via QR codes. 

Finally, a sixth zone comprised all elements of the exhibition that are not anchored in physical space: the online archive and its visualization by Kim Albrecht, radio performances regularly created by the participants: broadcasted, streamed, and offered as podcasts (Fig. \ref{fig:phone}).

Unlike in a typical exhibition, most projects and interfaces took shape after the initial opening; the website grew from a simple announcement into a sprawling resource that spilled over into radio platforms and telegram channels. While online and offline, synchronous and asynchronous components are simultaneously present, the spatial structure outlined above is designed with the goal to maximize the opportunity for encounters, interactions, and engagement across multiple channels.

\section{Infrastructuring as continuous development}
A diversity of channels and interconnections between physical and informational spaces can create logistic challenges—too much to keep track of and maintain. Spam and vandalism were a concern given the controversy around the “Querdenker” protest movement of pandemic deniers, which emerged during the time of the exhibition and manifested itself in public demonstrations, overpainted distancing markers, graffiti, and online trolling. While we welcomed critical reflections about the lockdown and its policy measures, we also wanted to avoid inadvertently creating a platform for misinformation and conspiracy theories. This effort required continuous maintenance and care, especially in the Telegram channel associated with the project. 

\begin{figure}
    \begin{subfigure}{0.5\linewidth}
        \centering
        \includegraphics[height=2.1in]{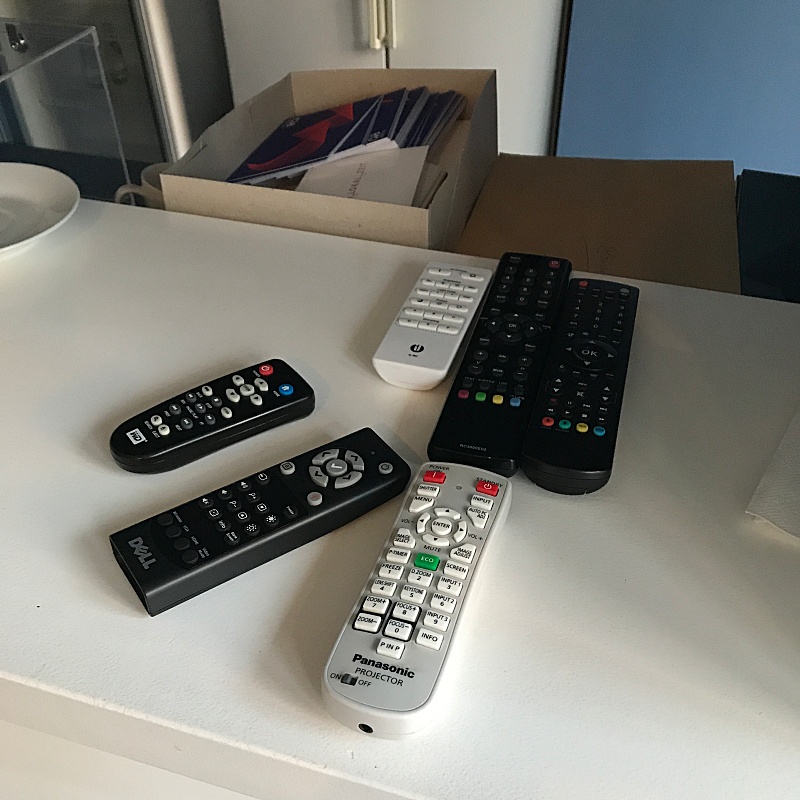}
        \caption{Georg Spehr's performance "Let me quickly go and turn on the gallery," Feb 21.}
        \label{fig:turn}
    \end{subfigure}
    ~
    \begin{subfigure}{0.5\linewidth}
        \centering
        \includegraphics[height=2.1in]{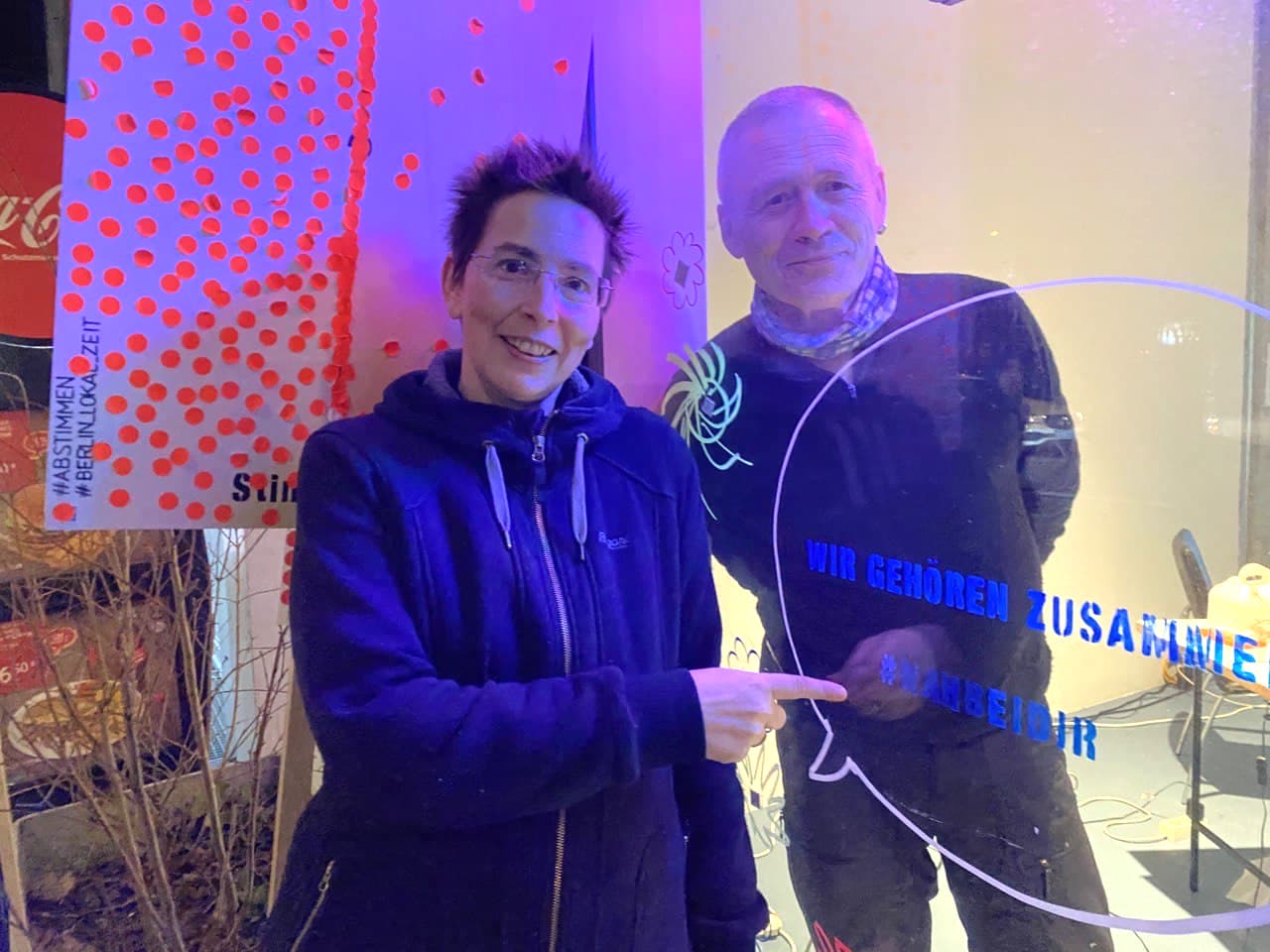}
        \caption{The "close to you" installation in its second version, katrinem and Sam Auinger, Feb 21.}
    \label{fig:close}
    \end{subfigure}%
    \caption{Maintenance and continuous revisions}
\end{figure}

The same is true for the physical aspects of the project. Throughout the exhibition, artist Max Joy repeatedly re-arranged and expanded his LED matrix displays showing a continuously updated list of COVID-related neologisms. He and artist katrinem regularly interacted with pedestrians as they spent time in the home office installation. Susanne Jaschko, who had to repeatedly redraw the chalk circles of “close to you,” responded to these frequent interactions by moving the circles to the glass facade, encouraging pedestrians to interact with the artists inside the gallery and post selfies on the Telegram channel. Other tasks of maintenance are more mundane (Fig. \ref{fig:close}). Artist Georg Spehr, who lives close to the space, took on the daily responsibility of switching on the devices in the gallery and documented this maintenance performance in a radio episode (Fig. \ref{fig:turn}).

The QR code facade, perhaps the most direct link between the online and the physical realm, was also continuously rearranged and updated by the participants. QR codes, two-dimensional matrix barcodes that can encode text or URLs, have been explored as an urban interaction modality since the early 2000s.\footnote{See for example the 2005 project Semapedia, which used QR codes to connect physical places with their respective wikipedia entry: \url{https://web.archive.org/web/20051228024514/http://www.semapedia.org/}} However, since their use was often cumbersome and required proprietary smartphone apps, they never became widely popular; their bad reputation also led to some controversy in the group. The pandemic, however, has finally led to their widespread adoption, as restaurants and shops used them for online menus and transactions to avoid physical contact. Decisive for the late success was that the smartphone operating systems IOS and Android have recently started offering native QR code recognition whenever a smartphone’s camera is activated. Our “media facade” consisted of a QR code matrix, continuously updated, re-arranged, and manually annotated with an easily removable chalk pen. Since every aspect of the project was represented online—from installations to performances and the archive of observations—this physical installation was crucial for offering flexible and serendipitous access to materials while liberating us from the rigid navigational structure of a website. As resources were accessible in their relevant physical context, physical visitors to the gallery did not have to traverse online menus for reaching them. 

\begin{figure}
\begin{subfigure}{0.49\linewidth}
    \centering
    \includegraphics[height=1.9in]{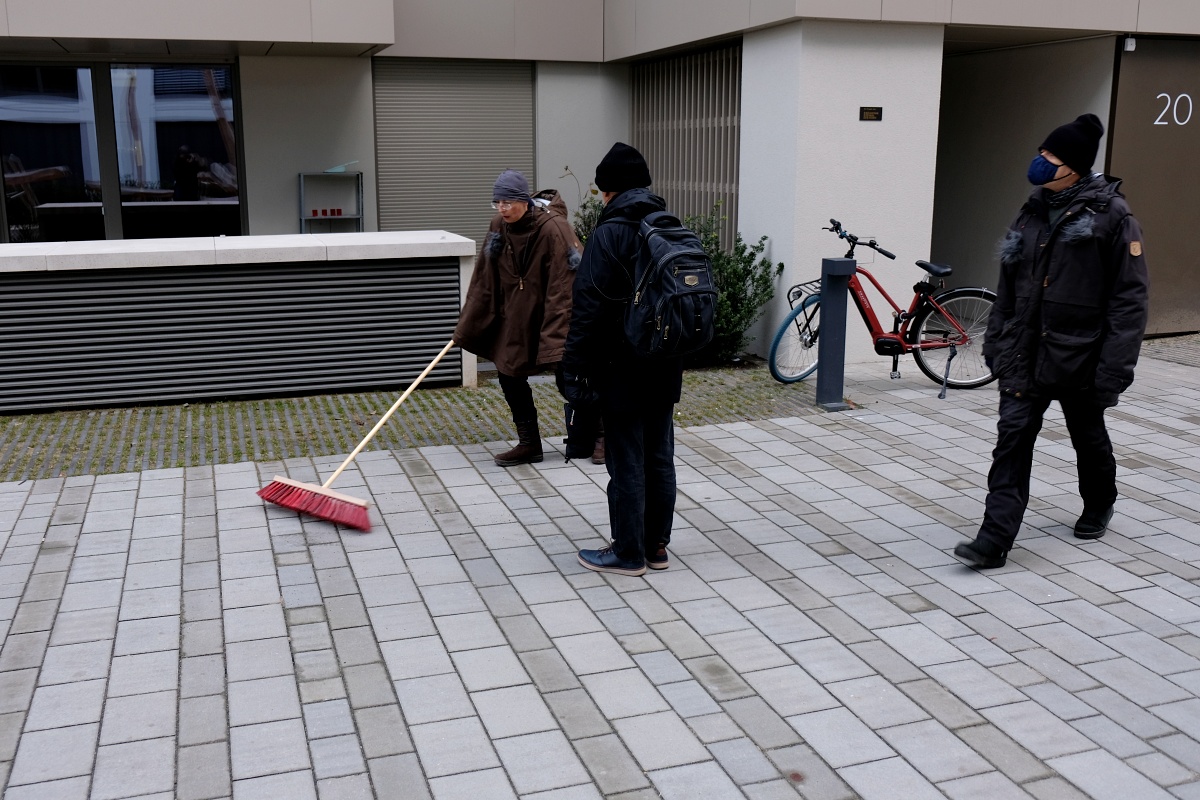}
    \caption{Soundwalk katrinem, Sam Auinger, Peter Cusack, Luisenpark.}
    \label{fig:3mic1}
\end{subfigure}
~
\begin{subfigure}{0.49\linewidth}
    \centering
    \includegraphics[height=1.9in]{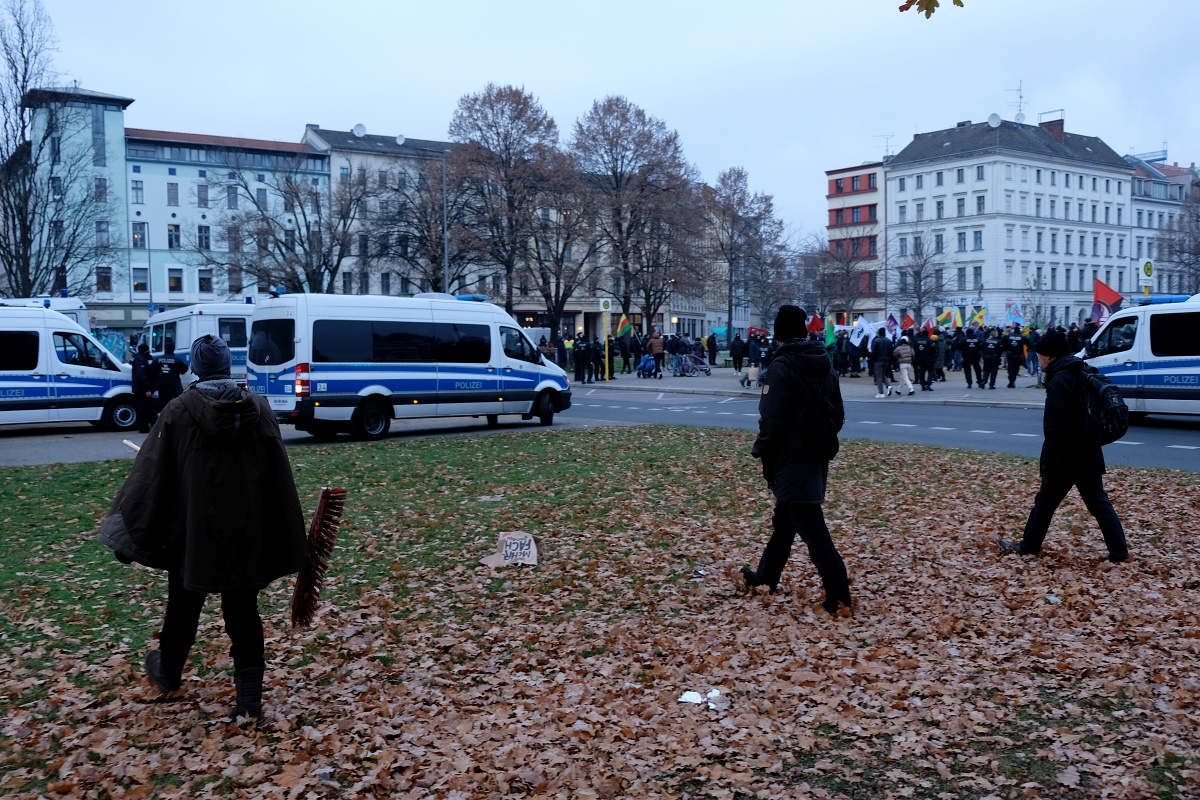}
    \caption{Soundwalk at the site of an anti-corona demonstration.}
    \label{fig:3mic2}
\end{subfigure}
\caption{Example of a small group performance offered as part of the project, a soundwalk exploring the sounds of the pandemic city, from quiet backyards to a demonstration of corona skeptics.}
\label{fig:soundwalk}
\end{figure}

To manage the complexity of our hyperlocal information space requires adjusting the toolbox; certain technologies and techniques are better suited to an improvisational approach than others. Consequently, focus of planning in the group shifted from defining the exhibition and the individual tasks for its production to planning for ad-hoc creation, growth, and change. While we used the QR codes mainly as a means for information access, the various mechanisms for bidirectional interaction with the public were kept deliberately analog— through the participation in soundwalks and their individual reenactment, toolkits for creating experiments and performances in public space, the interactions with artists in the gallery, or the collection of physical votes (Fig. \ref{fig:soundwalk}). 

Explicitly planning for an improvisational approach required finding alternatives for every aspect of the exhibition that would require decisions made weeks in advance. This includes the curation of the exhibition, its material aesthetics, the choice and design of interactive technologies, and finally, the communication strategies for introducing the project to the public. In all of these efforts, the interfaces between physical and informational spaces are not governed by a fixed infrastructure but by infrastructures that arise from ongoing activities. Whereas traditional media facades circumscribe the domain of possible content and interactions through its intrinsic specifications, in our case, meaning is created through extrinsic relations between people, things, and places. The space of mediated interactions is continuously created, extended, and modified by the participants.

\section{An improvisation based approach media architecture}
Beyond the pragmatic requirements of co-production, an improvisational approach has also epistemological implications for conceptualizing media architecture. First of all, it challenges a number of tenacious dichotomies: between the roles of designers/developers and users, the design of infrastructures and their use, between researchers and their research subjects. As noted earlier, the infrastructure is traditionally considered permanent, while the participants are treated as interchangeable. This separation is also reflected in the academic practices of evaluation: a cohort of N individuals are randomly selected to evaluate the design intervention according to a fixed protocol. Approaches such as participatory action research (PAR) reject such models and their claims of generalizability, involving communities in the research design and giving them voice in how the results are published and used \cite{kemmis2013}. The concept of the participant itself is, of course, not uncontroversial either since it is often used as a euphemism glossing over different roles, intentions, and degrees of control \cite{arnstein1969,cooke2001}. In BERLIN\_LOKAL\_ZEIT, the participants are authors, co-creators, and decision makers about almost every aspect of the project. They are not randomly sampled and interchangeable but form a group with established relationships that grew stronger during the course of the project. 

The infrastructures and technologies have undergone several shifts over the course of the project. Early on in the project, the physical media installations in the gallery were considered the heart of the project. This emphasis gradually shifted to performances and self-directed experiments as the project progressed. Towards the end of the project, the radio programs became a new center of the project, as the participants started to create documentaries, interviews, and podcasts about the local community. The QR-code facade was initially created as an alternative form to access the website, but turned into a bulletin board for new events and contributions as the project went on. 

Innovations that emerged during the show include the \textit{home office} project, which allowed the artist sitting at the desk how people interact with parts of the exhibition, according to katrinem, who proposed this component. As Max Joy points out, the communication with pedestrians became central—everyone tried to interact with the people in the gallery, knock on the window, waved. In an interview, Max drew parallels to the \textit{Fluxus} art movement, which followed to goal to establish art and art-making as a part of everyday life \cite{smith1998}. In this respect, the mediated channels were not just useful for connecting with the public, as originally intended, but became increasingly important to facilitate discourse inside the group, as Sven Sappelt, director of the CLB gallery observed. The initially central Slack workspace, created for the pandemic archive, moved to the background and was overshadowed by the Aporee radio platform developed by participant Udo Noll. Also the Telegram channel developed into a new center, serving both as an interface to the public as well as an internal organization tool. Participants found it liberating to lose the strict formats for communication, as sound artist Sam Auinger pointed out. The discourse among the group facilitated discourse with the public. 

The exhibition, initially planned for six weeks, ended up running for more than three months. Initially, the improvisational approach created some anxiety. Nobody knew what would happen, and if something would happen, as Sven Sappelt recalls. But this anxiety quickly gave way to an ongoing process of research, discourse, and creation. As Ursula Rogg notes, there was no endpoint to the project—the exhibition never stopped because a group within the group to keep the collaboration going. Some artists received special grants from the city that allowed them to maintain their artistic practice, but they were in the minority among the participants. The group reallocated resources for the exhibition to support those in the group who were struggling. Max Joy emphasizes the restorative function of the project - it creates support (another platform metaphor), creates energy and new connections. The engagement with other participants was experienced as deeply meaningful, the daily maintenance procedures of the exhibition space ("like taking care of our little baby," as katrinem put it) resulted in more interpersonal contacts, even with social distancing rules in place. 

Some participating artists, whose everyday routines were locked into tight production schedules before the pandemic, realigned their professional self-image through the project. As sound artist Sam Auinger recalls, usually after spending two months with a project, everyone is exhausted. Here, a different rhythm emerged, turned into an artistic way of life. Nika Radić also noted that this project allowed her to re-evaluate what it means to work as an artist. Sound artist Peter Cusack asked, can I enjoy something because something bad happened? 

\section{Conclusion}
We cannot claim that the insights from this experiment are generalizable, and we think such an ambition would be inappropriate. As Sven Sappelt notes, every individual experiences the pandemic differently depending on their personal circumstances. However, the project points to new strategies. As Sappelt continues, the responses of the cultural sector to the pandemic tend to fall into two alternatives: online streaming or closing down. With its improvisational response to the lockdown policies, BERLIN\_LOKAL\_ZEIT developed creative offerings that differ from what other institutions have proposed. Online components were designed to enhance the flexibility under lockdown conditions: walks into urban space could be conducted alone, in pairs, or small groups, guided by instructions and radio programs. Whether an interface to the public became more prominent in the project was a result of how the interface related to its context. Nothing was abstract—every contribution was connected to objects and places that could be touched, even as touching was considered suspicious. 

In a time when many people are locked into back-to-back online meetings, the project has a fundamentally restorative focus, and keep the physical public space of the city attractive, prevent to lose the kids to the digital world, as Sven Sappelt put it. He himself re-evaluated the time supposedly lost moving between meetings in different locations prior to the pandemic, which he now considers a form of positive friction, a time to contemplate and regenerate. The interfaces of BERLIN\_LOKAL\_ZEIT, in their analog  configurations, aim to similarly provide positive friction, act against the compression of time experienced by Sappelt and many others as the gaps between virtual appointments disappear. 

With regards to media architecture, this paper makes two points. First, we propose a perspective that considers urban media interfaces as improvisation-driven infrastructures (or \textit{improstructures}). In this perspective, technological arrangements and structures are always considered provisional and subject to revision, while the underlying human relationships are emphasized. This is the opposite of the idea of an generally applicable solution that offers a stable structure that works for an interchangeable population of users.  Second, it suggests that the characteristic aspect of media architecture should not be sought in the integration of media technology into the structure of a building but in the strong coupling of information and mediated discourse with a particular place. As such, the project underlines the importance of place-based media architecture during a time when the expression, contact, and discourse in public space are severely curtailed. Our case study details how discourse can unfold in physical space using analog and digital means, taking place simultaneously in multiple channels. There is not a single interface to accomplish this task; new channels are introduced and explored, while others move into the background. Associating media with a concrete place goes beyond GPS location tags, but means embodying them into a concrete action, object, or performance in physical space. Such an embodiment does not render them static, which is exemplified by the different temporalities explored through the project in the form of radio transmission, performances, and physical interactions. BERLIN\_LOKAL\_ZEIT programmatically describes the setting of a narrative that is continued and re-written by each intervention of a participant or the audience. 

\begin{acks}
The exhibition was a team effort and we want to thank all artists and participants (in alphabethical order): Kim Albrecht, Ingrid Beirer, Peter Cusack, Eliot Felde, Maren Hartmann, Martina Huber, Almut Hüfler, Susanne Jaschko, Max Joy, katrinem, Udo Noll, Nika Radić, Ursula Rogg, Sven Sappelt, Holger Schulze, Paul Scraton, Georg Spehr, Zoe Spehr, Hannes Strobl, Linh Hoang Thuy.
\end{acks}

\bibliographystyle{ACM-Reference-Format}
\bibliography{Zotero.bib}


\begin{thebibliography}{19}


\ifx \showCODEN    \undefined \def \showCODEN     #1{\unskip}     \fi
\ifx \showDOI      \undefined \def \showDOI       #1{#1}\fi
\ifx \showISBNx    \undefined \def \showISBNx     #1{\unskip}     \fi
\ifx \showISBNxiii \undefined \def \showISBNxiii  #1{\unskip}     \fi
\ifx \showISSN     \undefined \def \showISSN      #1{\unskip}     \fi
\ifx \showLCCN     \undefined \def \showLCCN      #1{\unskip}     \fi
\ifx \shownote     \undefined \def \shownote      #1{#1}          \fi
\ifx \showarticletitle \undefined \def \showarticletitle #1{#1}   \fi
\ifx \showURL      \undefined \def \showURL       {\relax}        \fi
\providecommand\bibfield[2]{#2}
\providecommand\bibinfo[2]{#2}
\providecommand\natexlab[1]{#1}
\providecommand\showeprint[2][]{arXiv:#2}

\bibitem[\protect\citeauthoryear{Arnstein}{Arnstein}{1969}]%
        {arnstein1969}
\bibfield{author}{\bibinfo{person}{Sherry~R. Arnstein}.}
  \bibinfo{year}{1969}\natexlab{}.
\newblock \showarticletitle{A ladder of citizen participation}.
\newblock \bibinfo{journal}{\emph{Journal of the American Institute of
  planners}} \bibinfo{volume}{35}, \bibinfo{number}{4} (\bibinfo{year}{1969}),
  \bibinfo{pages}{216--224}.
\newblock
\urldef\tempurl%
\url{http://www.tandfonline.com/doi/abs/10.1080/01944366908977225}
\showURL{%
\tempurl}


\bibitem[\protect\citeauthoryear{Caldwell and Foth}{Caldwell and Foth}{2014}]%
        {caldwell2014}
\bibfield{author}{\bibinfo{person}{Glenda~Amayo Caldwell} {and}
  \bibinfo{person}{Marcus Foth}.} \bibinfo{year}{2014}\natexlab{}.
\newblock \showarticletitle{{DIY} media architecture: open and participatory
  approaches to community engagement}. In \bibinfo{booktitle}{\emph{Proceedings
  of the 2nd {Media} {Architecture} {Biennale} {Conference}: {World} {Cities}}}
  \emph{(\bibinfo{series}{{MAB} '14})}. \bibinfo{publisher}{Association for
  Computing Machinery}, \bibinfo{address}{Aarhus, Denmark},
  \bibinfo{pages}{1--10}.
\newblock
\showISBNx{978-1-4503-3302-3}
\urldef\tempurl%
\url{https://doi.org/10/ghtfkf}
\showDOI{\tempurl}


\bibitem[\protect\citeauthoryear{Caldwell and Foth}{Caldwell and Foth}{2017}]%
        {caldwell2017}
\bibfield{author}{\bibinfo{person}{Glenda~Amayo Caldwell} {and}
  \bibinfo{person}{Marcus Foth}.} \bibinfo{year}{2017}\natexlab{}.
\newblock \showarticletitle{{DIY}/{DIWO} media architecture: {The}
  {InstaBooth}}.
\newblock \bibinfo{journal}{\emph{Using information and media as construction
  material}} (\bibinfo{year}{2017}), \bibinfo{pages}{61--80}.
\newblock


\bibitem[\protect\citeauthoryear{Chang}{Chang}{2016}]%
        {chang2016}
\bibfield{author}{\bibinfo{person}{Heewon Chang}.}
  \bibinfo{year}{2016}\natexlab{}.
\newblock \bibinfo{booktitle}{\emph{Autoethnography as {Method}}}.
\newblock \bibinfo{publisher}{Routledge}.
\newblock
\showISBNx{978-1-315-43336-3}
\newblock
\shownote{Google-Books-ID: vXCTDAAAQBAJ.}


\bibitem[\protect\citeauthoryear{Ciborra}{Ciborra}{1996}]%
        {ciborra1996}
\bibfield{author}{\bibinfo{person}{Claudio~U. Ciborra}.}
  \bibinfo{year}{1996}\natexlab{}.
\newblock \showarticletitle{The {Platform} {Organization}: {Recombining}
  {Strategies}, {Structures}, and {Surprises}}.
\newblock \bibinfo{journal}{\emph{Organization Science}} \bibinfo{volume}{7},
  \bibinfo{number}{2} (\bibinfo{year}{1996}), \bibinfo{pages}{103--118}.
\newblock
\showISSN{1047-7039}
\urldef\tempurl%
\url{http://www.jstor.org/stable/2634975}
\showURL{%
\tempurl}


\bibitem[\protect\citeauthoryear{Cooke and Kothari}{Cooke and Kothari}{2001}]%
        {cooke2001}
\bibfield{author}{\bibinfo{person}{Bill Cooke} {and} \bibinfo{person}{Uma
  Kothari}.} \bibinfo{year}{2001}\natexlab{}.
\newblock \bibinfo{booktitle}{\emph{Participation: the {New} {Tyranny}?}}
\newblock \bibinfo{publisher}{Zed Books}.
\newblock
\showISBNx{978-1-85649-794-7}


\bibitem[\protect\citeauthoryear{Cunha, Cunha, and Kamoche}{Cunha
  et~al\mbox{.}}{1999}]%
        {cunha1999}
\bibfield{author}{\bibinfo{person}{Miguel Pina~e Cunha}, \bibinfo{person}{Joao
  Vieira~da Cunha}, {and} \bibinfo{person}{Ken Kamoche}.}
  \bibinfo{year}{1999}\natexlab{}.
\newblock \showarticletitle{Organizational {Improvisation}: {What}, {When},
  {How} and {Why}}.
\newblock \bibinfo{journal}{\emph{International Journal of Management Reviews}}
  \bibinfo{volume}{1}, \bibinfo{number}{3} (\bibinfo{date}{Sept.}
  \bibinfo{year}{1999}), \bibinfo{pages}{299--341}.
\newblock
\showISSN{1460-8545, 1468-2370}
\urldef\tempurl%
\url{https://doi.org/10.1111/1468-2370.00017}
\showDOI{\tempurl}


\bibitem[\protect\citeauthoryear{Dantec and DiSalvo}{Dantec and
  DiSalvo}{2013}]%
        {dantec2013}
\bibfield{author}{\bibinfo{person}{Christopher A~Le Dantec} {and}
  \bibinfo{person}{Carl DiSalvo}.} \bibinfo{year}{2013}\natexlab{}.
\newblock \showarticletitle{Infrastructuring and the formation of publics in
  participatory design}.
\newblock \bibinfo{journal}{\emph{Social Studies of Science}}
  \bibinfo{volume}{43}, \bibinfo{number}{2} (\bibinfo{date}{April}
  \bibinfo{year}{2013}), \bibinfo{pages}{241--264}.
\newblock
\showISSN{0306-3127, 1460-3659}
\urldef\tempurl%
\url{https://doi.org/10.1177/0306312712471581}
\showDOI{\tempurl}


\bibitem[\protect\citeauthoryear{Egyedi and Mehos}{Egyedi and Mehos}{2012}]%
        {egyedi2012}
\bibfield{author}{\bibinfo{person}{Tineke~M. Egyedi} {and}
  \bibinfo{person}{Donna~C. Mehos}.} \bibinfo{year}{2012}\natexlab{}.
\newblock \bibinfo{booktitle}{\emph{Inverse {Infrastructures}: {Disrupting}
  {Networks} from {Below}}}.
\newblock \bibinfo{publisher}{Edward Elgar Publishing},
  \bibinfo{address}{Cheltenham, UK}.
\newblock
\showISBNx{978-1-84980-301-4}


\bibitem[\protect\citeauthoryear{Haeusler}{Haeusler}{2009}]%
        {haeusler2009}
\bibfield{author}{\bibinfo{person}{M.~Hank Haeusler}.}
  \bibinfo{year}{2009}\natexlab{}.
\newblock \bibinfo{booktitle}{\emph{Media {Facades}: {History}, {Technology},
  {Content}}}.
\newblock \bibinfo{publisher}{Avedition}.
\newblock
\showISBNx{978-3-89986-107-5}
\newblock
\shownote{Google-Books-ID: 9z3iOwAACAAJ.}


\bibitem[\protect\citeauthoryear{Hughes}{Hughes}{1987}]%
        {hughes1987}
\bibfield{author}{\bibinfo{person}{Thomas~Parke Hughes}.}
  \bibinfo{year}{1987}\natexlab{}.
\newblock \showarticletitle{The evolution of large technological systems}.
\newblock \bibinfo{journal}{\emph{The political economy of science, technology,
  and innovation}} (\bibinfo{year}{1987}), \bibinfo{pages}{51--82}.
\newblock


\bibitem[\protect\citeauthoryear{Kemmis, McTaggart, and Nixon}{Kemmis
  et~al\mbox{.}}{2013}]%
        {kemmis2013}
\bibfield{author}{\bibinfo{person}{Stephen Kemmis}, \bibinfo{person}{Robin
  McTaggart}, {and} \bibinfo{person}{Rhonda Nixon}.}
  \bibinfo{year}{2013}\natexlab{}.
\newblock \bibinfo{booktitle}{\emph{The {Action} {Research} {Planner}: {Doing}
  {Critical} {Participatory} {Action} {Research}}}.
\newblock \bibinfo{publisher}{Springer Science \& Business Media}.
\newblock
\showISBNx{978-981-4560-67-2}
\newblock
\shownote{Google-Books-ID: GB3IBAAAQBAJ.}


\bibitem[\protect\citeauthoryear{Kloeckl}{Kloeckl}{2020}]%
        {kloeckl2020}
\bibfield{author}{\bibinfo{person}{Kristian Kloeckl}.}
  \bibinfo{year}{2020}\natexlab{}.
\newblock \bibinfo{booktitle}{\emph{The {Urban} {Improvise}:
  {Improvisation}-{Based} {Design} for {Hybrid} {Cities}}}.
\newblock \bibinfo{publisher}{Yale University Press}.
\newblock


\bibitem[\protect\citeauthoryear{Lucero}{Lucero}{2018}]%
        {lucero2018}
\bibfield{author}{\bibinfo{person}{Andrés Lucero}.}
  \bibinfo{year}{2018}\natexlab{}.
\newblock \showarticletitle{Living {Without} a {Mobile} {Phone}: {An}
  {Autoethnography}}. In \bibinfo{booktitle}{\emph{Proceedings of the 2018
  {Designing} {Interactive} {Systems} {Conference}}}. \bibinfo{publisher}{ACM},
  \bibinfo{address}{Hong Kong China}, \bibinfo{pages}{765--776}.
\newblock
\showISBNx{978-1-4503-5198-0}
\urldef\tempurl%
\url{https://doi.org/10/gh5x7z}
\showDOI{\tempurl}


\bibitem[\protect\citeauthoryear{Offenhuber and Schechtner}{Offenhuber and
  Schechtner}{2018}]%
        {offenhuber2018a}
\bibfield{author}{\bibinfo{person}{Dietmar Offenhuber} {and}
  \bibinfo{person}{Katja Schechtner}.} \bibinfo{year}{2018}\natexlab{}.
\newblock \showarticletitle{Improstructure - an improvisational perspective on
  smart infrastructure governance}.
\newblock \bibinfo{journal}{\emph{Cities}}  \bibinfo{volume}{72}
  (\bibinfo{date}{Feb.} \bibinfo{year}{2018}), \bibinfo{pages}{329--338}.
\newblock
\showISSN{0264-2751}
\urldef\tempurl%
\url{https://doi.org/10.1016/j.cities.2017.09.017}
\showDOI{\tempurl}


\bibitem[\protect\citeauthoryear{Offenhuber and Seitinger}{Offenhuber and
  Seitinger}{2014}]%
        {offenhuber2014b}
\bibfield{author}{\bibinfo{person}{Dietmar Offenhuber} {and}
  \bibinfo{person}{Susanne Seitinger}.} \bibinfo{year}{2014}\natexlab{}.
\newblock \showarticletitle{Over the rainbow: information design for
  low-resolution urban displays}. In \bibinfo{booktitle}{\emph{Proceedings of
  the 2nd media architecture biennale conference: {World} cities}}.
  \bibinfo{publisher}{ACM}, \bibinfo{pages}{40--47}.
\newblock


\bibitem[\protect\citeauthoryear{Smith}{Smith}{1998}]%
        {smith1998}
\bibfield{author}{\bibinfo{person}{Owen~F. Smith}.}
  \bibinfo{year}{1998}\natexlab{}.
\newblock \bibinfo{booktitle}{\emph{Fluxus: {The} {History} of an {Attitude}}}.
\newblock \bibinfo{publisher}{San Diego State University Press},
  \bibinfo{address}{San Diego}.
\newblock
\showISBNx{1-879691-51-5}


\bibitem[\protect\citeauthoryear{Wiethoff and Hussmann}{Wiethoff and
  Hussmann}{2017}]%
        {wiethoff2017b}
\bibfield{author}{\bibinfo{person}{Alexander Wiethoff} {and}
  \bibinfo{person}{Heinrich Hussmann}.} \bibinfo{year}{2017}\natexlab{}.
\newblock \bibinfo{booktitle}{\emph{Media {Architecture}: {Using} {Information}
  and {Media} as {Construction} {Material}}}.
\newblock \bibinfo{publisher}{Walter de Gruyter GmbH \& Co KG}.
\newblock
\showISBNx{978-3-11-045387-4}
\newblock
\shownote{tex.ids= wiethoff2017 googlebooksid: FaY2DgAAQBAJ.}


\bibitem[\protect\citeauthoryear{Wouters, Claes, and Moere}{Wouters
  et~al\mbox{.}}{2018}]%
        {wouters2018a}
\bibfield{author}{\bibinfo{person}{Niels Wouters}, \bibinfo{person}{Sandy
  Claes}, {and} \bibinfo{person}{Andrew~Vande Moere}.}
  \bibinfo{year}{2018}\natexlab{}.
\newblock \showarticletitle{Hyperlocal {Media} {Architecture}: {Displaying}
  {Societal} {Narratives} in {Contested} {Spaces}}. In
  \bibinfo{booktitle}{\emph{Proceedings of the 4th {Media} {Architecture}
  {Biennale} {Conference}}} \emph{(\bibinfo{series}{{MAB18}})}.
  \bibinfo{publisher}{Association for Computing Machinery},
  \bibinfo{address}{New York, NY, USA}, \bibinfo{pages}{76--83}.
\newblock
\showISBNx{978-1-4503-6478-2}
\urldef\tempurl%
\url{https://doi.org/10.1145/3284389.3284490}
\showDOI{\tempurl}


\end{thebibliography}
\end{document}